\documentclass{cimento}
\usepackage{graphicx}
\title{Survival analysis of the Swift Optical Transient data}
\author{L.G.~Bal\'azs\from{i1}\ETC,
I.~Horv\'ath\from{i2}, Zs.~Bagoly\from{i3}, \atque
A.~M\'esz\'aros\from{i4}\from{i5}}
 \instlist{\inst{i1}Konkoly Observatory,
Budapest, Hungary
          \inst{i2} Bolyai Military University, Budapest, Hungary
          \inst{i3} E\"otv\"os University, Budapest, Hungary
          \inst{i4} Charles University, Prague, Czech Republic;
          \inst{i5} Stockholm Observatory, Sweden}
\PACSes{\PACSit{00.00}{By the way, which PACS is it, the 00.00?
GOK.}
\PACSit{---.---}{\ldots}}

\begin{document}
\maketitle
\begin{abstract}In a systematic search of the OTs at GRBs the
Swift satellite determined only an upper limit of the apparent
brightness in a significant fraction of cases. Combining these
upper limits with the really measured OT brightness we obtained a
sample well suited to survival analysis. Performing a Kaplan-Meier
product limit estimation we obtained an unbiased cumulative
distribution of the $V$ visual brightness. The  $\log_{10}(N(V))$
logarithmic cumulative distribution can be well fitted with a
linear function of $V$ in the form of $\log_{10}(N(V))= 0.234\,V +
const$. We studied the dependence of $V$ on the gamma ray
properties of the bursts. We tested the dependence on the fluence,
$T_{90}$ duration and peak flux. We found a dependence on the peak
flux on the 99.7\% significance level.
\end{abstract}

\section{Introduction}
Studying the optical transients (OT) of GRBs is one of the major
tasks of the Swift satellite. However, a significant fraction of
GRBs does not show detectable OT fluences. It means in the number
of cases one measured only an upper limit instead of a real event.
Treating only those bursts exceeding the limit of detection with a
high certainty one never can  be sure how representative is the
sample obtained in this way for the whole population of the GRBs
detected by the BAT.  Survival analysis offers a solution to
overcome this difficulty.

\section{Statistics with censored data}
If one knows a lower/upper bound for the measured data instead of
concrete values they are considered as censored. In mathematical
terms let we have  two independent stochastic  variables: $[X,Y]$.
If $X>Y$ then $X$ is detected, otherwise $Y$ which is an upper
bound for $X$ in this way. A sample representing these
observations consists of a mixture of real  and censored (upper
bound) data. In our case $X$ represents the brightness of an OT
 and  $Y$ an upper bound  for the non real (censored) detection. Kaplan and
Meier~\cite{kame} showed it is possible to estimate the true
distribution of a sample even in the case of censorship (for the
astronomical context see Feigelson and Nelson~\cite{fene}).

\section{Mathematical formulation}
According to Kaplan and Meier  the following estimate gives the
survival function:

\begin{equation}
S(x) = 1 - F(x) = \prod \limits_{i=1}^n P_i
\end{equation}

\noindent where  $F(x)$  is is the probability distribution
function and

\begin{equation}
P_i = \left[1- \frac{1}{n-i+1}\right]^{\delta_i}
\end{equation}

\noindent $n$ means the number of data, $\delta$ equals 1 at non
censored data and 0 in the censored case. It means the estimated
survival function jumps at real data and remains constant in the
case of censoring.

\section{Survival analysis with OT data}

\begin{figure}
\center
\includegraphics[width=0.6\linewidth]{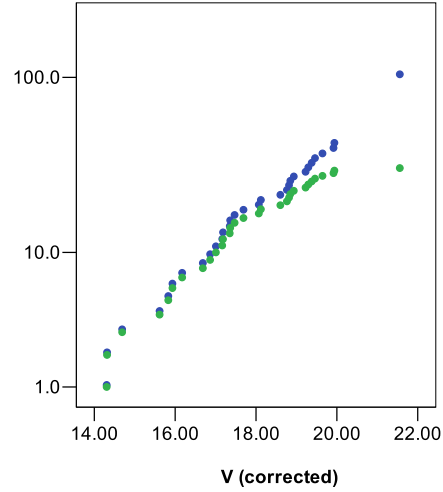}
\caption{Cumulative distribution of the observed events (green
dots) and that of obtained from the survival analysis (blue dots)
} \label{surv1}
\end{figure}

Until preparing this work Swift observed 144  events and detected
OT  in 31 cases. 73 bursts have only upper limits and the rest has
no optical data. Before making the survival analysis the effect of
the foreground extinction has to be removed. We used  the data of
Schlegel et al.~\cite{schleg}. Performing the analysis we derived
the true cumulative distribution of the V visual magnitudes which
can be compared with that of the measured optical  events, as
displayed in Figure \ref{surv1}.

The logarithmic cumulative distribution of the V magnitudes
obtained from the survival analysis can be  well fitted with the
following linear relationship:

\begin{equation}
\log_{10}N(V) = 0.234 \,V + const
\end{equation}

\noindent It is worth mentioning that a homogeneous spatial
distribution in Euclidean space would resulted in a coefficient of
0.6 instead of 0.234, as obtained. This result gives an important
constraint on estimating the spatial distribution of OT events.

\begin{figure}
\center
\includegraphics[width=0.6\linewidth]{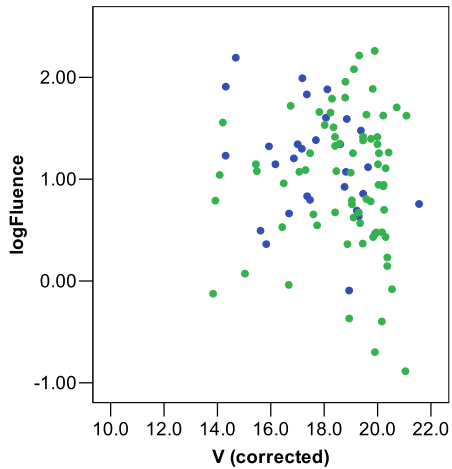}
\caption{Scatter plot between the extinction corrected V visual
brightness and the logarithmic peak flux. The $V$ magnitude of the
real events (blue dots) show a significant correlation on the
99.7\% level which is not the case with the censored data (green
dots)} \label{surv2}
\end{figure}

 The BAT on the Swift satellite  measured the
fluence, peak flux and $T_{90}$ duration in the 15-150 keV energy
range. We computed the Pearson linear correlation between the
extinction-corrected $V$ visual magnitude of the OTs and the
logarithms of quantities mentioned. We obtained significant
correlation only with the logarithmic peak flux at the 99.7\%
level (see Fig. \ref{surv2}).

\acknowledgments This research was supported by  OTKA grant
T048870 and by a grant from Wenner-Gren Foundation (A.M.)

\end{document}